\documentstyle[aps,prb,multicol,epsf]{revtex}
\draft
\begin{document}
\newcommand{\cid}{c^{{\scriptscriptstyle \dag}}_{i\sigma}}
\newcommand{\cjd}{c^{{\scriptscriptstyle \dag}}_{j\sigma}}
\newcommand{\cidbar}{c^{{\scriptscriptstyle \dag}}_{i\bar{\sigma}}}
\newcommand{\cibar}{c^{}_{i\bar{\sigma}}}
\newcommand{\ci}{c^{}_{i\sigma}}
\newcommand{\cj}{c^{}_{j\sigma}}
\newcommand{\ed}{\epsilon^{d}_{x}}
\newcommand{\es}{\epsilon^{s}_{x}}
\newcommand{\sss}{\scriptscriptstyle}
\newcommand{\k}{{\bf k}}
\newcommand{\G}{{\rm G}}
\newcommand{\tp}{t^{\prime}}
\newcommand{\yy}{{\displaystyle \sum_{yy^{\prime}}}}
\newcommand{\yyts}{{\textstyle \sum_{yy^{\prime}}}}
\newcommand{\yp}{y^{\prime}}
\newcommand{\minplus}{{\sss -+}}
\title{Local moment formation in zinc doped cuprates.}
\author{B.C. den Hertog and M.P. Das}  
\address{\small{{\it Department of Theoretical Physics,}} \\
\small{{\it  Research School of Physical
Sciences and Engineering,}} \\
\small{{\it The Australian National University, Canberra ACT, 
0200  Australia.}}}

\maketitle                                                
\begin{abstract}
We suggest that when zinc is substituted for copper in the copper oxide planes
 of high $T_{c}$ superconductors, it does not necessarily have a valency of
 $2^{{\scriptscriptstyle +}}$. Rather, the  valency of a zinc impurity should be   determined by its surrounding medium. In order to  study this hypothesis, we  examine the effect of static impurities 
inducing  diagonal  disorder  within a one band Hubbard
 model coupled to a localised state.
 We use this model to discuss the physics of zinc doping in the cuprates. 
Specifically, we discuss the formation of local moments near impurity sites
 and the modification of the  transverse spin susceptibility in the
 antiferromagnetic state.
\end{abstract}
\vspace*{1cm}
\begin{multicols}{2}
\section{Introduction}
The use of a substitutional impurity such as Zn for Cu within the copper oxide
 planes of the high $T_{c}$ materials\cite{Xiao} is considerably helpful for 
gaining insight into both the normal and superconducting properties of these 
systems. Zinc is known to produce a universal rate of $T_{c}$ depression in
La,Y,Bi and Tl based high temperature superconductors\cite{Agarwal},  lending 
support to a universal pairing mechanism in these cuprates. It is also known to
 severely affect  normal state electronic transport properties, 
\cite{Zagoulaev1,Uchida} for example inducing semiconducting behaviour in otherwise
 metallic
 cuprates.\cite{Agarwal,Das} The magnetic behaviour of the high $T_{c}$ oxides
 is also modified in the presence of zinc. Antiferromagnetic order is 
 destroyed \cite{Xiao2,Keimer} (albeit not as rapidly as superconductivity) and there is a dramatic effect on dynamical spin 
fluctuations. Whilst the underdoped cuprates are known to possess a spin pseudo gap, Kakurai {\it et.al.}\cite{Kakurai} have shown that unlike pure
 ${\rm YBa_{2}Cu_{3}O_{6.6}}$, zinc doped ${\rm YBa_{2}Cu_{2.9}Zn_{0.1}O_{6.6}}
$ has no gap like structure in the spin fluctuation spectrum, on the contrary  there is a significant enhancement in spectral weight at low frequencies. 

It has been suggested that these normal state effects and the observed reduction in $T_{c}$
 are related to the  formation of local moments  observed in
 the vicinity of the zinc dopants.\cite{Williams,Cooper,Alloul1,Alloul2} These moments can be detected by NMR 
techniques \cite{Alloul1,Alloul2,Walstedt,Ishida} and contribute a Curie law 
paramagnetism to the magnetic susceptibility.\cite{Jee,Zagoulaev2}
 The purpose of this paper is to discuss from a theoretical viewpoint, the
 formation of such local moments and how they relate to antiferromagnetism and
 superconductivity.

 Recently there has been a  considerable  amount of interest in local moment 
formation and its effects on magnetic and superconducting behaviour. Yet there 
are relatively few theoretical works\cite{Ziegler,Basu,Sen,Basu2,Ovchinnikov} which  
attempt to show microscopically how such local moments may form in a 
two dimensional strongly correlated fermion system. To date, it has been generally believed that zinc
 substitution results in a Zn$^{{\scriptscriptstyle ++}}$ ion situated at a
 copper site in the superconducting planes. This zinc ion is  in a closed $3d^{10}$ configuration and hence 
the $3d$ holes of the Cu lattice scatter off a static impurity of, from their
 perspective, very high energy. We wish to suggest here that one cannot assume
 automatically that the zinc atom is in this valence state. The valency of an
 atom is effected by its surrounding medium. Accordingly, it is not clear that
 the role of the zinc
  $4s$ orbital can be immediately discarded from the conceptual framework, particularly when  $3d$ fermions which are
experiencing  strong Coulomb repulsion  are present.
 In fact, Hao {\it et. al.}\cite{Hao} have shown in an electronic structure
 calculation of ${\rm La_{1.85}Sr_{0.15}Cu_{1-x}Zn_{x}O_{4}}$, that the partial 
 density of states of the $4s$ orbital is peaked right at the Fermi energy. 
More ab-initio calculations are required to produce a clearer picture on this point. 

In this work we consider local moment formation in a model which reflects the fact that the
zinc $3d$ electrons are tightly bound to the nucleus (absence of holes) whilst
 allowing a coupling between the local $4s$ orbital and the itinerant  states 
of the Cu planar network. To this end  our model possesses   both Hubbard and 
Anderson type qualities. We  show that, depending on various coupling 
strengths, local moments are able to form  within  the Hubbard gap and above 
and below the Hubbard bands, with spectral weight on and near to the impurity
 sites. Further, we  show that the transverse spin susceptibility at the
 antiferromagnetic wave vector ${\bf Q}=(\pi,\pi)$ (the lattice spacing is set to unity) indeed gains spectral weight at low frequencies due to the presence of such local moments.
\section{Theory}
 Our model Hamiltonian is the following: 
\begin{equation}
\label{model}
 H=H^{0} + H^{I} + H^{II} \; ,
\end{equation}
where
\[
\begin{array}{ccl}
 H^{0}&=&{\displaystyle \sum_{<ij>\sigma}}t(\cid\cj + h.c.) +
    U{\displaystyle \sum_{i}}\cid\ci\cidbar\cibar  \\
&&\\
&-&{\displaystyle \sum_{i\sigma}}\mu\cid\ci + {\displaystyle \sum_{x\sigma}}
(\es-\mu) s^{{\sss \dag}}_{x\sigma}s^{}_{x\sigma} \; , \\
& & \\
 H^{I}&=&{\displaystyle \sum_{x\sigma}}\ed c^{{\sss \dag}}_{x\sigma}c^{}_{x\sigma}\;,
 \\
& & \\
H^{II}&=&
{\displaystyle \sum_{<xy>\sigma}}t^{\prime}(s^{{\sss \dag}}_{x\sigma}c^{}_{y\sigma} +     h.c.) \; .
\end{array}
\]
The creation operator $\cid$ creates a hole in a $3d$ copper orbital at 
 lattice site $i$ with spin $\sigma$ whilst $\cj$ destroys a $3d$ hole with
 spin $\sigma$ at lattice site $j$. The first term in $ H^{0}$
 represents nearest neighbour hopping on a square lattice and the second term
 is the usual Hubbard interaction between $3d$ holes at the same site. The
 last term is $H^{0}$ is simply  the $4s$ orbital energy level where 
 $\es$ is
 the energy difference between the  copper $3d$ and zinc  $4s$ hole orbitals.
 The 
creation operator $s^{\sss \dagger}_{x\sigma}$ creates a hole in a zinc $4s$ 
orbital with spin $\sigma$. The
 index $x$ labels the impurity sites while $y$ labels an impurity site's first
 nearest neighbours. The energy 
$\ed$ is the  difference between copper and zinc $3d$  hole states. 
Here $\ed>0$ whilst 
$\es<0$ and $\mu$ is the chemical potential of the $3d$ system. Thus $H^{I}$ represents
 static impurity scattering of $3d$ holes off the zinc
 sites whilst $H^{II}$ reflects a coupling between the localised $4s$
 states and the continuum.

In two dimensions the Hubbard model itself has proven to be  rather difficult  when it comes to theoretical analysis. Consequently,
 the presence of extra terms in our initial Hamiltonian  makes our analytical task no
 easier. To this end, we employ the finite $U$ slave boson method of Kotliar
 and Ruckenstein\cite{Kotliar} (KR) which, at the mean field level, is in
 reasonable agreement with Monte Carlo data on the one band  2D Hubbard
 model\cite{Lilly}, and has had similar success with 
the three band Hubbard model.\cite{Emerya,Littlewood,den Hertog2,den
 Hertog1,Zhang} We use this technique to generate the antiferromagnetic
 long range order  which the 2D Hubbard model  is believed to possess  at half filling.\cite{Hirsch} We are then able to  
 consider the effect of the one body terms $ H^{I}$ and $H^{II}$.
 While our treatment is not self-consistent in the sense the we do not use
 the effect of $H^{I}$ and $H^{II}$ to recalculate the 
various bosonic parameters of the KR method, a Hartree Fock self consistent 
study of a similar model shows that corrections to this approximation are of higher order only.\cite{Basu2}

To generate the antiferromagnetic host we introduce four
 auxiliary boson fields representing each possible configuration of the $3d$ orbital at a particular site.\cite{Kotliar} That is, a $3d$ orbital may be empty, singly occupied with a spin $\sigma$ or doubly occupied.
 The fermionic creation operator is mapped $\cid\mapsto z^{}_{i\sigma}\cid$,
 where $z^{}_{i\sigma}$ is defined as
\begin{equation}
z^{}_{i\sigma}\equiv
\frac{e^{}_{i}p^{{\sss \dag}}_{i\sigma} + p^{}_{i\bar{\sigma}}d^{{\sss \dag}}_{i} }
{\sqrt{(1 - d^{{\sss \dag}}_{i}d^{}_{i} - p^{{\sss \dag}}_{i\sigma}p^{}_{i\sigma})
(1 - e^{{\sss \dag}}_{i}e^{}_{i} -
 p^{{\sss \dag}}_{i\bar{\sigma}}p^{}_{i\bar{\sigma}}) } } \; .
\end{equation}
There is  a conjugate mapping for the fermionic destruction operator. 
The boson fields $e^{}_{i},p^{}_{i\sigma}$ and $d^{}_{i}$ represent empty,
 single with spin $\sigma$ an double occupation of the $3d$ orbital at a site
 $i$ respectively.
The denominator in Eq. (2) ensures that the mapping becomes trivial in the limit $U\rightarrow 0$.
Unphysical states arising from the enlargened Hilbert space are eliminated via the following constraints:
\begin{equation}
\cid\ci=p^{{\sss \dag}}_{i\sigma}p^{}_{i\sigma} + d^{{\sss \dag}}_{i}d^{}_{i}
\end{equation}
and
\begin{equation}
e^{{\sss \dag}}_{i}e^{}_{i} + \sum_{\sigma}p^{{\sss \dag}}_{i\sigma}p^{}_{i\sigma} +
d^{{\sss \dag}}_{i}d^{}_{i} =1
\end{equation}
which imply charge conservation and completeness of the bosonic operators 
respectively. Introducing the constraints into the Hamiltonian $H^{0}$
 via Lagrange multipliers we have,
\begin{eqnarray}
H^{0}&=&\sum_{<ij>\sigma}t(q^{}_{\sigma}\cid\cj + h.c.) +
    U\sum_{i}d^{{\sss \dag}}_{i}d^{}_{i} \nonumber \\
&+&\sum_{i\sigma}(\lambda^{}_{i\sigma}- \mu)(\cid\ci - 
p^{{\sss \dag}}_{i\sigma}p^{}_{i\sigma} + d^{{\sss \dag}}_{i}d^{}_{i}) 
\nonumber \\
&+&\sum_{i}\lambda^{\prime}_{i}(e^{{\sss \dag}}_{i}e^{}_{i} + \sum_{\sigma}p^{{\sss \dag}}_{i\sigma}p^{}_{i\sigma} +d^{{\sss \dag}}_{i}d^{}_{i} - 1) 
\nonumber\\
&+&{\displaystyle \sum_{x\sigma}}\es
 s^{{\sss \dag}}_{x\sigma}s^{}_{x\sigma} \;, \nonumber \\
\end{eqnarray}
where $q_{ij\sigma}=z^{}_{i\sigma}z^{*}_{j\sigma}$.

To accommodate the antiferromagnetic phase of the Hubbard model we split the
 square lattice into two sub-lattices $A$ and $B$. The symmetry of this phase
 implies
\[
n_{{\sss A}\sigma} = n_{{\sss B}\bar{\sigma}}\equiv n_{\sigma} \; .
\]
 In terms of the
 bosons, 
\[
\langle p^{{\sss \dag}}_{{\sss A}\sigma}p^{}_{{\sss A}\sigma}\rangle= 
\langle p^{{\sss \dag}}_{{\sss B}\bar{\sigma}}p^{}_{{\sss B}\bar{\sigma}}\rangle\equiv
\langle p^{{\sss \dag}}_{\sigma}p^{}_{\sigma}\rangle  \; ,
\]
\[
\langle e^{{\sss \dag}}_{{\sss A}}e^{}_{{\sss A}}\rangle=\langle e^{{\sss \dag}}_{{\sss B}}e^{}_{{\sss B}}
\rangle\equiv\langle e^{{\sss \dag}}e^{}\rangle \; ,
\]
 and
\[
\langle d^{{\sss \dag}}_{{\sss A}}d^{}_{{\sss A}}\rangle=\langle d^{{\sss \dag}}_{{\sss B}} d^{}_{{\sss B}}\rangle \equiv
\langle d^{{\sss \dag}}d^{}\rangle \; .
\]
 We also have 
$\lambda^{}_{{\sss A}\sigma}=\lambda^{}_{{\sss B}\bar{\sigma}}\equiv \lambda^{}_{\sigma}$.
The partition function corresponding to $H^{0}$ can be calculated exactly in 
the fermion sector and  at the saddle point in the boson sector
(where the bosonic fields  become $c$ numbers).  The self consistent mean 
field equations can then be solved. At half filling the chemical potential 
$\mu=(\lambda_{{\sss \uparrow}} + \lambda_{{\sss \downarrow}})/2$.

At the mean field level, the quasiparticle bandstructure is given by
\begin{equation}
E_{\k\nu\sigma}= \nu\sqrt{\Delta^2 + q^2 \epsilon(\k)^2 }
\end{equation}
where $\epsilon(\k)$ is the 2D square lattice tight binding band structure,
 $\Delta=(\lambda_{{\sss \uparrow}} - \lambda_{{\sss \downarrow}})/2$ and
 $\nu=\pm$ which refers to the upper and lower Hubbard bands respectively. The
 quasiparticle amplitudes on each sub-lattice are:
\begin{equation}
a_{\k\nu\sigma}^2=\frac{q^2 \epsilon(\k)^2}{q^2 \epsilon(\k)^2 +
 (E_{\k\nu\sigma} - \sigma\Delta)^2} \; ,
\end{equation}
and by symmetry,
\begin{equation}
b_{\k\nu\bar{\sigma}}^2=a_{\k\nu\sigma}^2 \;,
\end{equation}
where $a_{\k\nu\sigma}$ refers to the $A$ sub-lattice and $b_{\k\nu\sigma}$ to the $B$ sub-lattice. 

The Green's function of $H^{0}$ may be written as,
\begin{equation}
{\rm{\bf G}}^{0}_{\sigma}({\bf k},\omega)=\left[
\begin{array}{ll}
{\rm G}^{0c}_{\sigma}({\bf k},\omega)& \\
& \\
& {\rm G}^{0s}_{\sigma}(\omega) \\
\end{array} \right] \; ,
\end{equation}
where ${\rm G}^{0c}_{\sigma}({\bf k},\omega)$ is the Green's function for the continuum states and is itself a ${\rm 2x2}$ matrix because of the sub-lattice 
structure. The Green's function 
$ {\rm G}^{0s}_{\sigma}(\omega)$ is due to the localised $4s$ state. Specifically, they are;
\begin{equation}
{\rm G}^{0c}_{\sigma}(\k,\omega)=\sum_{\nu}\left[
\begin{array}{cc}
a_{\k\nu\sigma}^2 & a_{\k\nu\sigma}^{*}b^{}_{\k\nu\sigma} \\
& \\
a^{}_{\k\nu\sigma}b_{\k\nu\sigma}^{*} & b_{\k\nu\sigma}^2 \\
\end{array} \right] \frac{1}{\omega - E_{\k\nu\sigma} \pm i\eta}
\end{equation}
and
\begin{equation}
{\rm G}^{0s}_{\sigma}(\omega)=\frac{1}{\omega  - (\es- \mu)}
\end{equation}
where $\pm i\eta$ refers to the upper and lower Hubbard bands respectively.

We  initially consider the effect of one zinc impurity positioned on the $A$ sub-lattice.  Formally it is simpler to first take into account the effect of
 $H^{I}$.
 The $T$- matrix expression for   the continuum Green's 
function in real space is:  
\begin{equation}
{\rm G}^{Ic}_{ij\sigma}(\omega)={\rm G}^{0c}_{ij\sigma}(\omega) + 
\frac{{\rm G}^{0c}_{ix\sigma}(\omega)\ed {\rm G}^{0c}_{xj\sigma}(\omega)}{1 -
 \ed {\rm G}^{0c}_{xx\sigma}(\omega)}
\; ,
\end{equation}
and of course $\G^{Is}_{\sigma}(\omega)=\G^{0s}_{\sigma}(\omega)$.

Including now  the scattering contributions of $H^{II}$, the full 
 continuum Green's function of $H$ in real space is:
\begin{equation}
\label{gc}
\G^{c}_{ij\sigma}(\omega)= \G^{Ic}_{ij\sigma}(\omega) + 
\frac{\yyts\G^{Ic}_{iy}(\omega)\tp\G^{Is}_{\sigma}(\omega)\tp 
\G^{Ic}_{\yp j\sigma}(\omega)}{1 - 
\tp\G^{Is}_{\sigma}(\omega)\tp\yyts\G^{Ic}_{y\yp\sigma}(\omega)} \; ,
\end{equation}
and the full local state Green's function can be written as
\begin{equation}
\label{g2}
\G^{s}_{\sigma}(\omega)=\frac{1}{w - (\es-\mu)  - \Sigma(\omega)}
 \; ,
\end{equation}
where $\Sigma(\omega)$ is  a self energy  given by
 $\Sigma(\omega)=\yyts\tp\G^{Ic}_{y\yp}(\omega)\tp$.

 The poles of $\G^{c}_{ij\sigma}(\omega)$ and $\G^{s}_{\sigma}(\omega)$ give
 the discrete states of the system. They are given by the
 solution of
\begin{equation}
\label{poles}
\begin{array}{ll}
\left(\omega - (\es-\mu)\right)\left(1-\ed\G^{0c}_{xx\sigma}(\omega)\right) - 
\yy {\tp}^{2}\ed\G^{0c}_{y\yp\sigma}(\omega) & \\
- \yy{\tp}^{2}\left(1 -\ed\G^{0c}_{xx\sigma}(\omega)\right)
 \G^{0c}_{xy\sigma}(\omega)
\G^{0c}_{\yp x\sigma}(\omega) = 0 \; .& \\
\end{array}
\end{equation}
Note that in the limit $\tp\rightarrow 0,\ed\rightarrow 0$ we recover the discrete $4s$ state. 

To determine if the discrete states (which are the solutions of
 Eq.(\ref{poles})) are localised in nature we can  calculate the
 probability of finding a particle with spin $\sigma$ occupying one of these states at the
 impurity site or at its nearest neighbours through the residues of the 
appropriate Green's functions. For example, for the
 $4s$ state we use Res$[\G^{s}_{\sigma}(\omega);\omega=E_{dis}]$, for the $3d$
 state at the $x$ site we have Res$[\G^{c}_{xx\sigma}(\omega);\omega=E_{dis}]$
 and for the nearest neighbour $3d$ states we use
 Res$[\G^{c}_{xy\sigma}(\omega);\omega=E_{dis}]$. Here $E_{dis}$ is the energy of the discrete state in question.

\section{Results}
\subsection{Local moments}
We have used in our calculations a value of  $U=6.4t$ 
 which produces a gap of $2\Delta=4t$ and a chemical potential for the half filled $3d$ antiferromagnetic system of $\mu=-3.49t$. We have also set the impurity $3d$ 
orbital energy at $\ed=9t$. This enables us to explore as a function of
 $\es-\mu$ and $\tp$ the nature of the discrete states which form in the whole
 system.

It is appropriate to first discuss the energy structure of the various orbital energy
 levels and quasiparticle bands. In Fig.1 we show this relationship. The 
copper $3d$ hole orbital energy is used as a reference. The two quasiparticle 
bands are shown as well as where the Fermi level sits and the quantity
 $\es - \mu$. Thus when $\es - \mu=0$ one is considering a zinc $4s$ hole
 level roughly $3.49 t$ below the copper $3d$ hole level. A large value of
 $|\es - \mu|$ would be favourable for zinc to acquire a valency of 
$2^{{\sss +}}$ as the $4s$ discrete state would become occupied by holes.

Solutions to Eq.(\ref{poles}) for  $\ed>0$ and for any values of $\es - \mu$ 
and $\tp$ 
 produce two discrete states above the upper Hubbard band (in the hole
 picture). With increasing $\ed$ they become completely localised to the 
impurity $3d$ orbital. Accordingly they represent localised zinc $3d$ electrons.
 The appearance of gap states or states below the lower band depends on the
 combination of the above  parameters.

\begin{figure}
\narrowtext
\epsfxsize=7cm
\epsfysize=5cm
\begin{center}
\hspace{1cm}\epsfbox{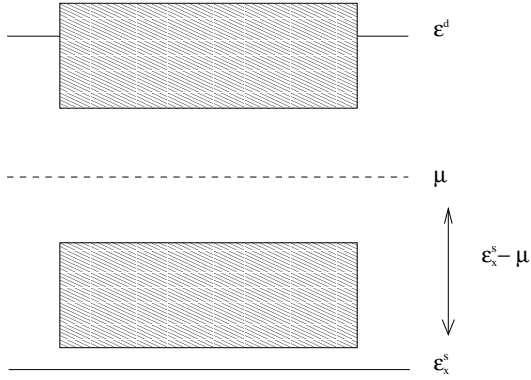}
\end{center}
\caption{Energy structure of the hole orbital energies and bands (shaded regions).}
\end{figure}

Without any coupling to the $4s$ state  the system produces one gap state
\cite{Sen} which has a reasonable amount of spectral weight
 on its nearest neighbours, producing a local moment. In the limit
 $\ed\rightarrow\infty$ the  spectral weight of the gap state is completely
 removed
 from the impurity site. In general, for fixed $\ed=9t$, a coupling to the
 $4s$ state produces one or two extra gap states depending on $\tp$ and
 $\es-\mu$. These are the remnants of the $4s$ localised states however some
 of their spectral weight has been shifted to the nearest neighbours by
 differing amounts depending on spin, which in effect produces a local moment
 occupying the $4s$ state. In Fig.2 we give an example of the graphical form
 of Eq.(\ref{poles}) for a parameter set of $\tp=0.25t$ and
 $\es - \mu= -2t$.
\vspace*{-5mm}
\begin{figure}
\narrowtext
\epsfxsize=8.6cm
\epsfysize=8.6cm
\begin{center}
\epsfbox{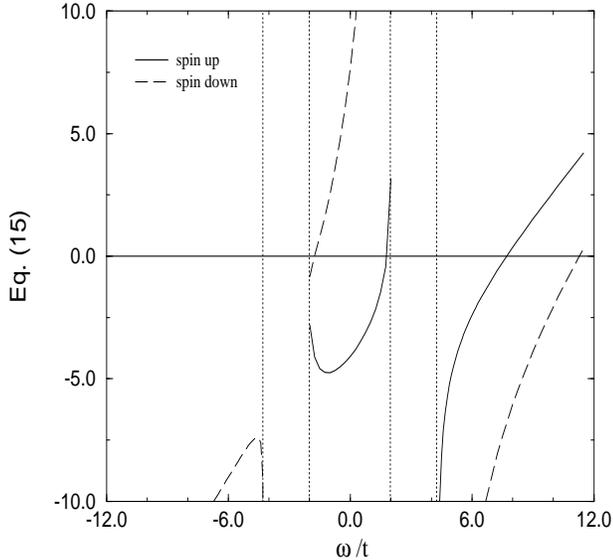}
\end{center}
\caption{The intersections with the horizontal axis give the discrete states
 of the system. Here we show an example for $\tp=0.25t,\es-\mu=-2t$ and
 $\ed=9t$.The region between the first two dotted vertical  lines is the
 lower band and the region between the second pair is the upper band.}
\end{figure}
We can see by the intersections with the horizontal axis in  Fig.2, that two gap states have formed as have two states above
 the hole upper band. These last two states are strongly localised in the zinc 
 $3d$ orbital and show that this orbital is filled with electrons. The up
 spin gap state has amplitude  mainly  on the zinc neighbours however the 
down spin gap state has amplitude  mainly in the zinc $4s$ orbital.
 Thus we have two local moment contributions.
 One situated at the zinc site and one spread mainly
 over the nearest neighbours. Both spins are in opposite alignment to the antiferromagnetically aligned  moments of their  respective sub-lattices. 

Both localised gap states are able to act as unpaired moments which should
 contribute a Curie like behaviour to the temperature dependent susceptibility.
 With more impurity doping and therefore more  localised states, the antiferromagnetic 
correlation length will be reduced leading to an eventual loss of 
antiferromagnetic order. Calculations by Sen {\it et al.}\cite{Sen} have shown
 that in the presence of local moments, the spin-wave  stiffness vanishes 
beyond a certain length scale, destroying long range antiferromagnetic order.   
\begin{figure}
\narrowtext
\epsfxsize=8.6cm
\epsfysize=8.6cm
\begin{center}
\epsfbox{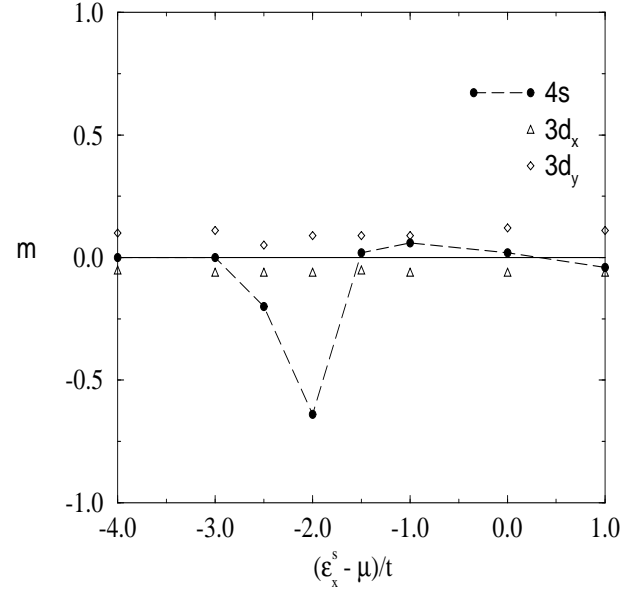}
\end{center}
\caption{Local moment $m=n^{}_{\uparrow} - n^{}_{\downarrow}$ as a function of
$(\es - \mu)/t$ for the 
$4s$ orbital, the $3d$ orbital at the zinc site$(3d_{x})$ and the $3d$ orbital of one of the nearest neighbours$(3d_{y})$.}
\end{figure}
For $\tp=0.25 t,$ if $\es-\mu$ is situated in the
 gap then   an up spin gap state and a down spin gap state form with large amplitudes in
 the $4s$ orbital and thus there is no local moment contribution from the $4s$
 states. Accordingly the zinc atom is nearly inert. An additional up spin gap state forms with amplitude mainly on the
 neighbouring sites producing a local moment. If however $\es - \mu$ is 
situated within a small energy window between $-3t$ and $-1.5t$, then the up
 spin gap state with 
amplitude mainly on the $4s$ orbital is lost, leaving only the down spin state which in effect leads to a 
 local moment at the zinc site. There is still a  local moment from
the remaining  up spin  gap state  whose amplitude is  found  mainly on neighbouring sites.  
If $\es - \mu$ is too negative then hole bound states are formed below the bands in the $4s$ orbitals. 
 In Fig.3 we show the net moment produced by the
 discrete states as a function of $\es -\mu$ in units of $t$ for $\tp=0.25 t$.

It can be seen  that the valence of a zinc 
impurity cannot  be simply assumed to be $2^{{\sss +}}$. Our purpose here has been to show   that the valence of a zinc atom can only be considered in the presence  of  strongly correlated $3d$ fermions.
\subsection{Spin Fluctuations}
It is instructive to examine the effect of localised states on   
spin fluctuation excitations. Attempts at calculating  the spin fluctuation  spectrum of
 an itinerant  antiferromagnet such as the 2D Hubbard model have generally
 involved a mean field plus RPA type approach\cite{Schrieffer} which attempts
 to incorporate Gaussian fluctuations around the  antiferromagnetic
 saddle point. This avenue
 of attack has been successful in producing  low frequency spin wave modes
 in the large $U$ limit with  dispersion  which agrees with linear spin wave 
theory of the 2D Heisenberg antiferromagnet.\cite{Kampf} Here, we do not attempt
to calculate a renormalised susceptibility by including fluctuations, but 
instead we calculate the modification of the bare susceptibility of
 the antiferromagnetic mean field state due to the presence of local moments
 in an attempt to illustrate that the presence of local moments causes an
 increase in spectral weight in the spin fluctuation spectrum, particularly at 
 low frequencies.

 Even at this level the  full calculation is in general quite complicated.
 Thus, as an example we calculate the contributions from localised states
 produced purely by the impurity energy $\ed$. Thus we set $\tp=0$ and so 
ignore any contributions from localised states formed by coupling of the 
conduction holes to the localised $4s$ state.

The transverse spin fluctuations of the rigid antiferromagnetic host  can be
 described by a 2x2 matrix
\begin{equation}
{\bf \chi}^{0c}({\bf Q},\omega)=\left[
\begin{array}{ll}
\chi_{aa}^{0c}({\bf Q},\omega)&\chi^{0c}_{ab}({\bf Q},\omega) \\
& \\
\chi^{0c}_{ba}({\bf Q},\omega)&\chi^{0c}_{bb}({\bf Q},\omega)
\end{array}\right]\; ,
\end{equation}
where $a$ and $b$ label the sub-lattices and the elements of
 $\chi^{0c}({\bf Q},
\omega)$ are defined as
\begin{equation}
\begin{array}{l}
\chi^{0c}_{mn}({\bf Q},\omega)=  \\
\\
{\displaystyle \frac{i}{2\pi N} 
 \sum_{{\bf p}}\int^{\infty}_{-\infty}dE^{\prime} \G^{ 0c}_{ mn{\sss \uparrow}}} 
({\bf p},E^{\prime})\G^{0c}_{nm{\sss \downarrow}}
({\bf p}- {\bf Q},E^{\prime}-\omega) \; , \\
\end{array}
\end{equation}
where $m$ and $n$ are either $a$ or $b$. The full spin susceptibility matrix elements are similarly defined. That is, we have
\begin{equation}
\begin{array}{l}
{\bf \chi}^{c}_{mn}({\bf Q},\omega)=  \\
\\
{\displaystyle \frac{i}{2\pi N} 
 \sum_{{\bf p}}\int^{\infty}_{-\infty}dE^{\prime}
 \G^{Ic}_{ mn{\sss \uparrow}}} 
({\bf p},E^{\prime})\G^{Ic}_{nm{\sss \downarrow}}
({\bf p}- {\bf Q},E^{\prime}-\omega) \; . \\
\end{array}
\end{equation}

For simplicity we assume a small concentration of impurities distributed evenly over the two sub-lattices such  that multiple scattering from different sites may be safely neglected. The diagonal continuum Green's functions can be written in momentum space as
\begin{eqnarray}
\G^{c}_{mm\sigma}({\bf k},\omega)&=&\G^{0c}_{mm\sigma}({\bf k},\omega) +
\frac{\ed n^{}_{x_{m}}}{1 - \ed \G^{0c}_{x_{m}x_{m}\sigma}}\G^{0c}_{mm}({\bf k}
,\omega)^{2} \nonumber \\
&+&\frac{\ed n^{}_{x_{\bar{m}}}}{1 - \ed \G^{0c}_{x_{\bar{m}}x_{\bar{m}}\sigma}}
\G^{0c}_{m\bar{m}\sigma}({\bf k},\omega)\G^{0c}_{\bar{m}m\sigma}({\bf k},\omega) \; , \nonumber \\
\end{eqnarray} 
where $n_{x_{m}}$ is the concentration of impurities on the $m$ sub-lattice.
The  diagonal components of the susceptibility can be calculated by carrying
 out  contour integrations in the upper and lower halves of the complex plane, making sure to 
take into account the poles on the real axis due to the discrete states. 
 
In Fig.4 we have evaluated at ${\bf Q}=(\pi,\pi)$ the imaginary part of the localised state contributions to  the diagonal
 components of ${\bf \chi}^{c}({\bf Q},\omega)$ up to first order in impurity
 concentration of $n_{x_{a}}=n_{x_{b}}=0.05$ (full curve).  We have calculated the  contributions to the spectral weight of spin fluctuations  by scattering processes within the bands and they are negligible. The dashed curves 
are the sum of
 the imaginary parts of the diagonal components of 
$\chi^{0c}({\bf Q},\omega)$. At this level they represent  particle-hole excitations only. We have set $\ed=5t$ which produces for each impurity one gap state and 
two states above the upper Hubbard band in the hole picture. It can be seen
 that the contribution from these localised states is an enhancement in the spectral weight at low frequencies.

 The increase in spectral weight at low frequencies would be further enhanced 
 if one were to include the off diagonal processes which contribute to 
$\chi^{c}({\bf Q},\omega)$ (which are due to inter sub lattice excitations), 
as well as the contribution from any extra localised states produced by a coupling of
 the itinerant states to the zinc $4s$ states via $\tp$. 
\begin{figure}
\narrowtext
\epsfxsize=8.6cm
\epsfysize=8.6cm
\begin{center}
\epsfbox{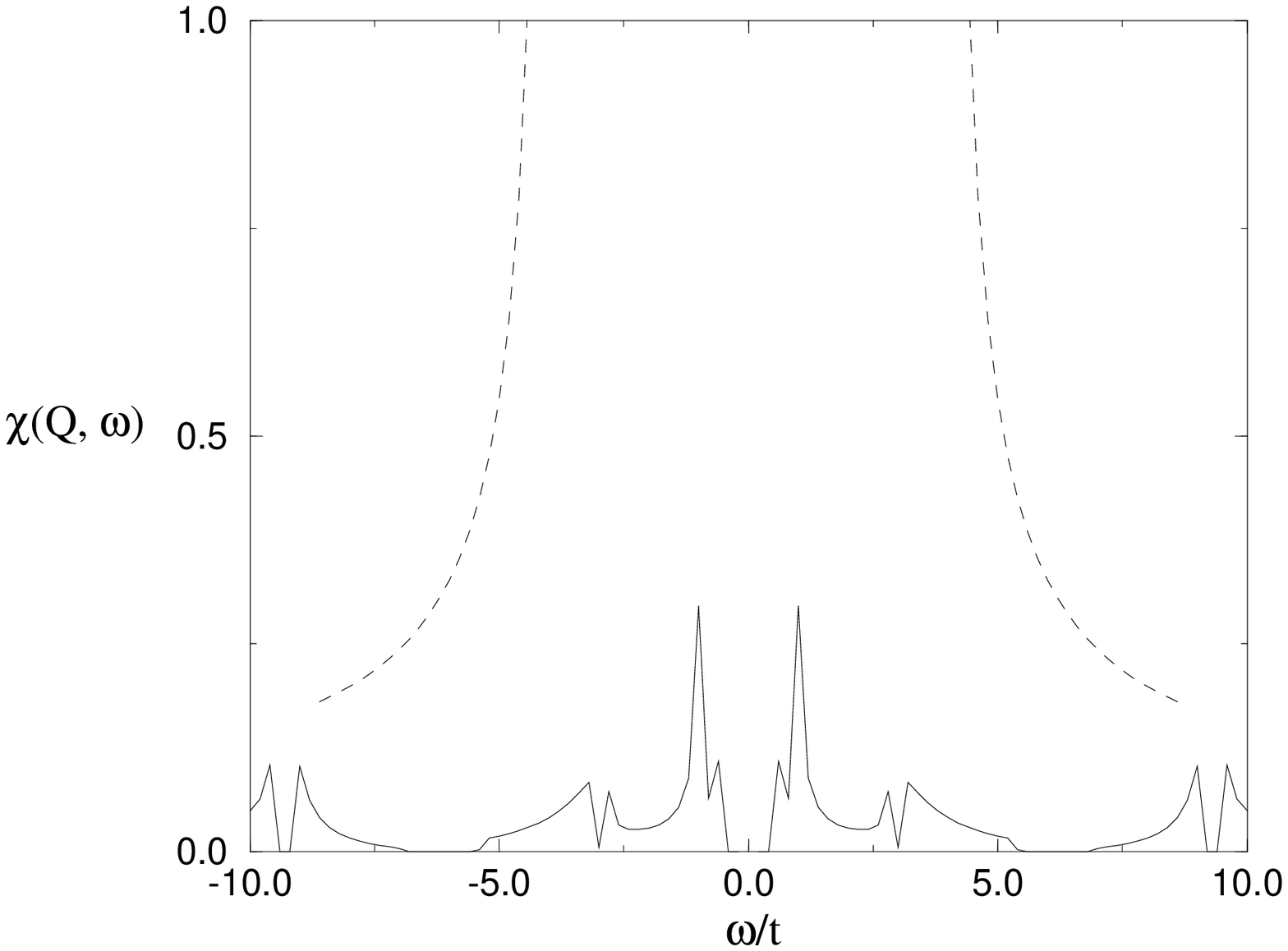}
\end{center}
\caption{Imaginary part of the transverse spin susceptibility at ${\bf Q}=(\pi,\pi)$. Here $\ed=5t$. The dashed line is the sum of the diagonal components of Im $\chi^{0c}({\bf Q},\omega)$ whilst the solid line represents the contributions from the discrete states.}
\end{figure}
Indeed if this increase in spectral weight is also produced away from half filling then this phenomena may go some way in explaining how the spin pseudo gap in the underdoped compounds vanishes with zinc doping. As Tallon {\it et al.}\cite{Tallon} have pointed out, for a small amount of zinc doping  the pseudo gap is suppressed near the vicinity of the  dopants and may be completely  suppressed when the mean spacing between zinc atoms falls below the pseudo gap correlation length.

In this picture superconductivity would also be  extinguished when  the
 pair correlation length falls below the mean spacing between zinc atoms. Pair
 breaking could occur via the Abrikosov-Gorkov mechanism due  to effective
 magnetic scattering, however there is evidence to suggest that this effect
 is too small to account for the rapid reduction of $T_{c}$.\cite{Walstedt} If the order parameter has $d_{x^2 - y^2}$ symmetry 
then in addition to the this,  strong potential scattering from the zinc impurities may be enough to suppress $T_c$ at a rate seen by experiment.

 Alternatively, if in the underdoped materials,  the formation of a spin
 pseudo gap is related to the formation of local singlet pairs, then  
suppression of this gap due to an increase in spin excitation spectral weight 
 may  also be a possible explanation for the reduction of $T_{c}$.

\section{Conclusions}
We have attempted to show that when  substituting  zinc for copper within the
copper oxide planes one cannot assume that zinc is in the $2^{{\sss +}}$ state. Rather
, its valency is sensitive to the presence of a strongly correlated system. We 
have shown that local moment formation is possible on and near the zinc sites 
and that they produce an enhancement of the  low energy spectral weight of 
 antiferromagnetic spin fluctuations.

\section{Acknowledgements}
We would like to thank T.C. Choy and M. Gulacsi for useful discussions. B.C.dH would like to thank the Commonwealth Government of Australia  for providing financial assistance.

\end{multicols}
\end{document}